\documentstyle[procl,epsfig]{article}

\def\be{\begin{equation}}
\def\ee{\end{equation}}
\def\bea{\begin{eqnarray}}
\def\eea{\end{eqnarray}}

\newcommand{\bq}{\begin{equation}}
\newcommand{\eq}{\end{equation}}

\newcommand{\ba}{\begin{eqnarray}}
\newcommand{\ea}{\end{eqnarray}}

\begin{document}

\begin{titlepage}

\begin{flushleft}
{\tt hep-ph/9607329} \\
July 1996 
\end{flushleft}

\setcounter{page}{0}

\mbox{}
\vspace*{\fill}
\begin{center}
{\Large\bf On Small-{\boldmath $x$} Resummations for the Evolution of}\\

\vspace{1.7mm}
{\Large \bf Unpolarized and Polarized}\\

\vspace{1.2mm}
{\Large\bf Non--Singlet and Singlet Structure Functions}\\

\vspace{3em}
\large
Johannes Bl\"umlein$^a$, Stephan Riemersma$^{a}$, and Andreas Vogt$^{b}$

\vspace{3em}
\normalsize
{\it $^a$DESY--Zeuthen} \\
{\it Platanenallee 6, D--15735 Zeuthen, Germany} \\
\vspace{4mm}
{\it $^b$Institut f\"ur Theoretische Physik, Universit\"at W\"urzburg}\\
{\it Am Hubland, D--97074 W\"urzburg, Germany} \\
\normalsize
\vspace*{\fill}
{\large \bf Abstract}
\end{center}
\vspace*{3mm}

\noindent
A brief survey is given of recent results on the resummation of 
leading small-$x$ terms for unpolarized and polarized non--singlet 
and singlet structure function evolution.

\vspace*{\fill}
\begin{center}
\large
{\it Contribution to the Proceedings of the International Workshop on}\\
{\it $\!\!\!\!\!$Deep Inelastic Scattering and Related Phenomena, Rome, 
April 1996$\!\!\!\!$}

\end{center}

\end{titlepage}

\title{
ON SMALL-{\boldmath $x$} RESUMMATIONS FOR THE EVOLUTION OF  
UNPOLARIZED AND POLARIZED NON--SINGLET AND SINGLET
STRUCTURE FUNCTIONS~\footnote{Talk presented by J. Bl\"umlein.}}

\author{ JOHANNES BL\"UMLEIN, STEPHAN RIEMERSMA }

\address{
DESY--Zeuthen, Platanenallee 6, D--15735 Zeuthen, Germany}

\author{ ANDREAS VOGT }

\address{
 Institut f\"ur Theoretische Physik, Universit\"at W\"urzburg, \\
 Am Hubland, D--97074 W\"urzburg, Germany}

\maketitle\abstracts{
A brief survey is given of recent results on the resummation of 
leading small-$x$ terms for unpolarized and polarized non--singlet 
and singlet structure function evolution.}

\section{Introduction}
\noindent
The evolution kernels of both non--singlet and singlet parton
densities contain large logarithmic contributions for small fractional
momenta $x$. In all--order resummations of these terms in the limit $x 
\rightarrow 0$, one naturally faces the problem of factorization and 
renormalization scheme dependence. Therefore these resummations have to 
be performed in the frame of the corresponding renormalization group 
equations. In the following we will discuss the resulting small-$x$ 
resummations for the anomalous dimensions relevant to the different DIS 
processes and their quantitative consequences.

For unpolarized deep-inelastic processes the leading small-$x$ 
contributions to the gluonic anomalous dimensions behave like~\cite{RE1} 
($N$ is the Mellin moment)
\begin{equation}
 \left ( \frac{\alpha_s}{N - 1} \right )^k \:\:\: \leftrightarrow
 \:\:\: \frac{1}{x} \,  \alpha_{s}^{k} \, \ln^{k-1} x \:\: .
\end{equation}
The corresponding quark anomalous dimensions, being one power down in 
$\ln x$, have been derived in ref.~\cite{RE2}. The leading terms of  
all anomalous dimensions for the non--singlet~\cite{RE3} and polarized 
singlet~\cite{RE4} evolutions are given by
\begin{equation}
 N \left (  \frac{\alpha_s}{N^2} \right )^k \:\:\: \leftrightarrow 
 \:\:\: \alpha_{s}^{k} \, \ln^{2k-2} x \:\: .
\end{equation}
The resummation of these terms can be completely derived by means of 
perturbative QCD. Its effect on the behaviour of the various DIS 
structure functions, however, is necessarily determined as well by the 
behaviour of the input parton densities at an initial scale $Q_0^2$, and
is therefore not predictable within perturbative QCD but has to be 
determined by experiment. Thus the resummation effect can only be studied
via the evolution of structure functions over some range in $Q^2$, which 
moreover probes the anomalous dimensions at all $ z \geq x $ via the 
Mellin convolution with the parton densities.

In leading (LO) and next--to--leading order (NLO) QCD the complete 
anomalous dimensions are known~\cite{RE5}. Hence the effect of the 
all--order resummation of the most singular parts of the splitting
functions as $x \rightarrow 0$ concerns only orders higher than 
$\alpha_{s}^2$.
Due to the Mellin convolution also terms less singular as $x \rightarrow
0$ may contribute substantially at these higher orders as well. In some 
cases the existence of such terms is enforced by conservation laws. 
For the non--singlet `-'-evolution fermion--number conservation implies
\begin{equation}
\int_0^1 \! dz \sum_{k=1}^{\infty} \alpha_s^k P_k^{-}(z) = 0 \:\: .
\end{equation}
Correspondingly, energy--momentum conservation holds for the unpolarized 
singlet evolution. Even in the polarized singlet case, where no 
conservation laws constrain the anomalous dimensions, the LO and NLO 
results exhibit terms which are less singular by one power in $N$ and 
have about the same coefficient but with opposite sign, cf.\ ref.~\cite
{RE9}. Since such contributions and further corrections are not yet 
known to all orders, it is reasonable to estimate their possible impact 
by corresponding modifications of the resummed anomalous dimensions 
$\Gamma(N,\alpha_s)$. Possible examples studied within refs.~$^{6-10}$ 
are:
\begin{equation}
\begin{array}{cl}
{\rm A:} & \Gamma(N, \alpha_s)
\rightarrow  \Gamma(N, \alpha_s) - \Gamma(1, \alpha_s)
\\
{\rm B:} & \Gamma(N, \alpha_s)
\rightarrow  \Gamma(N, \alpha_s)(1 - N)
\\
{\rm C:} & \Gamma(N, \alpha_s)
\rightarrow  \Gamma(N, \alpha_s)
(1 - 2N + N^2)
\\
{\rm D:} & \Gamma(N, \alpha_s)
\rightarrow  \Gamma(N, \alpha_s)(1 - 2N + N^3) \:\: ,
\end{array}
\end{equation}
where $ N \rightarrow N-1 $ for the case of eq.~(1).
Clearly the presently known resummed terms are only sufficient for 
understanding the small-$x$ evolution, if the difference of the results
obtained by these prescriptions are small.

\section{Resummation of dominant terms  for $x \rightarrow 0$}
\vspace{-2mm}
\subsection{Unpolarized non--singlet structure functions}
\noindent
The numerical effects due to the resummation of the $O(\alpha_s \ln^2x)$
terms (2) have been studied in refs.~\cite{RE7,RE8} for the structure 
functions $F_2^{\, ep} - F_2^{\, en}$ and $xF_3^{\, \nu N}(x,Q^2)$  
over a wide range of $x$ and $Q^2$. The resummed terms beyond NLO lead to
corrections on the level of 1\% and below even at extremely small $x$. 
$K$--factors of about 10 as claimed in ref.~\cite{RE12a} are not 
confirmed. Furthermore less singular terms can alter the resummation 
correction by a factor of about 3.

\vspace{-2mm}
\subsection{Unpolarized singlet structure functions}
\noindent
The quantitative impact of the resummation of the leading small-$x$ 
terms in the gluonic~\cite{RE1} and quarkonic anomalous dimensions~\cite
{RE2} has been studied for the sea quark ($S$) and gluon ($g$) 
distributions and the structure function $F_2^{ep}$ in refs.~\cite{RE6} 
and ref.~\cite{RE10}. The latter analysis confirms the results of the
former one. Related investigations were carried out in refs.~\cite
{RE6a}.
\begin{center}
\vspace*{-1mm}
\mbox{\epsfig{file=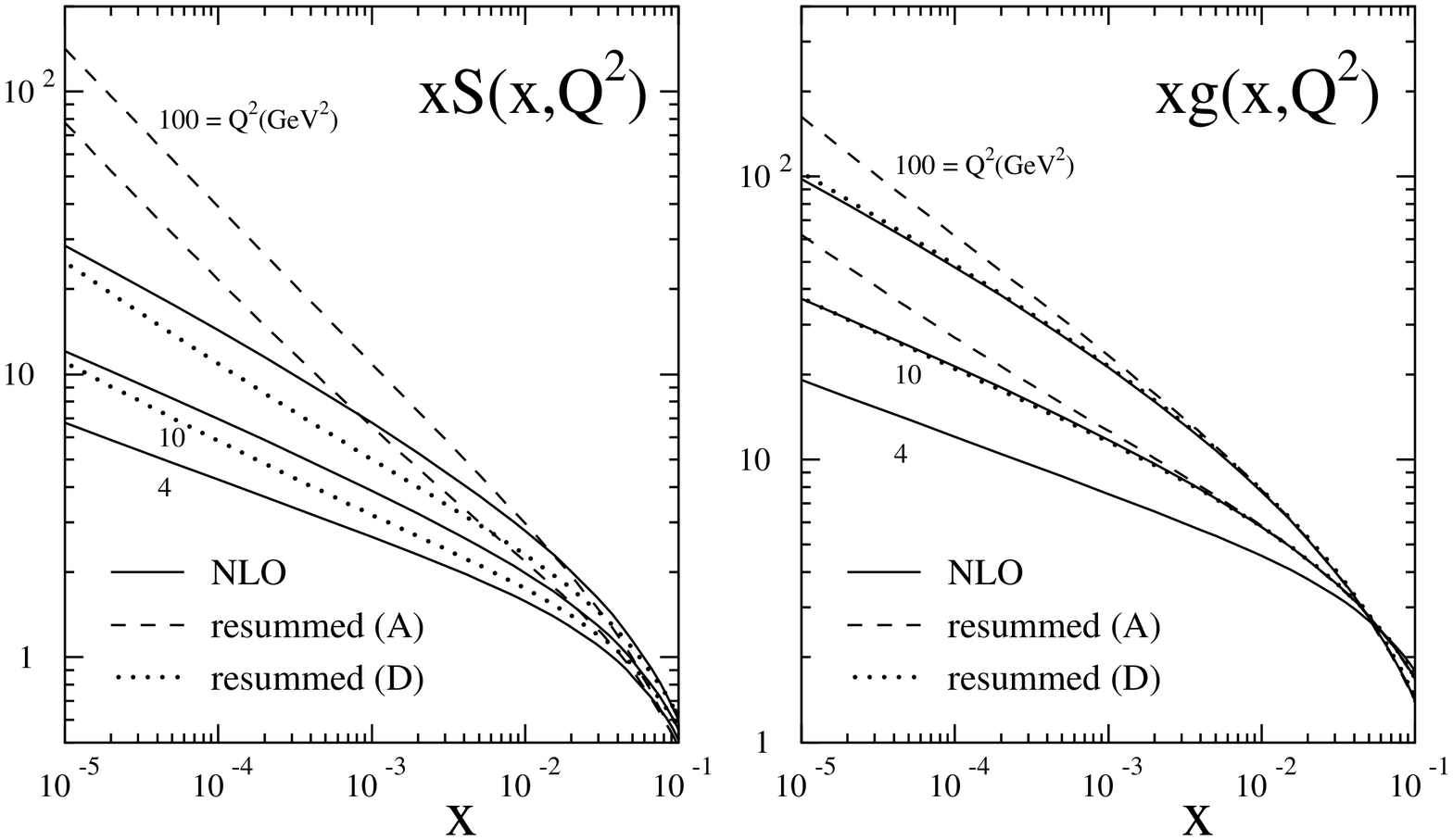,height=5.1cm,width=9.5cm}}
\vspace*{-4mm}
\end{center}
{\footnotesize
 \sf Figure~1:~The evolution of the gluon and total sea densities with 
 the resummed kernels~\cite{RE1,RE2} as compared to the NLO results. Two 
 different prescriptions for implementing the momentum sum rule have been
 applied, cf.\ eq.~(4). For details on initial parton densities etc see 
 ref.~\cite{RE10}.}
\vspace{2mm}

\noindent
In Fig.~1 the evolution of initial distribution $xS,\, xg \sim x^{-0.2}$
at $ Q_{0}^2 = 4 \mbox{ GeV}^2 $ is displayed. The effect of the 
resummation is very large. For the quarks and hence $F_2$ it is entirely 
dominated by the quarkonic pieces of ref.~\cite{RE2}. Two examples of 
additional less singular terms are shown. Their vital importance is 
obvious from the fact that the choice of (D) in eq.~(4) leads to even an 
overcompensation of the enhancement due to the leading small-$x$ terms.

\vspace{-2mm}
\subsection{Polarized non--singlet structure functions}
\noindent
This case was investigated in refs.~\cite{RE7,RE8} numerically for the
structure function combination $g_1^{ep}\! -g_1^{en}$ for two different 
parametrizations of the non-perturbative initial distributions. Results 
on the interference structure function $g_{5, \gamma Z}^{ep}(x,Q^2)$ 
(cf.~ref.~\cite{REBK}) can be found in ref.~\cite{RE10}. As in the 
unpolarized case the corrections obtained are of the order of 1\% with 
respect to the NLO results in the kinematical ranges experimentally
accessible in the foreseeable future. Huge $K$--factors of about 10 
or larger expected for this case in ref.~\cite{RE12b} are not present.
Again less singular terms in the anomalous 
dimensions are only marginally suppressed.

\vspace{-2mm}
\subsection{Polarized singlet structure functions}
\noindent
Resummation relations for amplitudes related to the singlet anomalous 
dimensions of polarized DIS have recently been given~\cite{RE4}. Explicit
analytical and numerical results for the evolution kernels beyond NLO
have been derived on this basis in ref.~\cite{RE9}, including an 
all-order symmetry relation among the elements of the anomalous dimension
matrix and a discussion of the supersymmetric case.
\begin{center}
\vspace*{-1.2mm}
\mbox{\epsfig{file=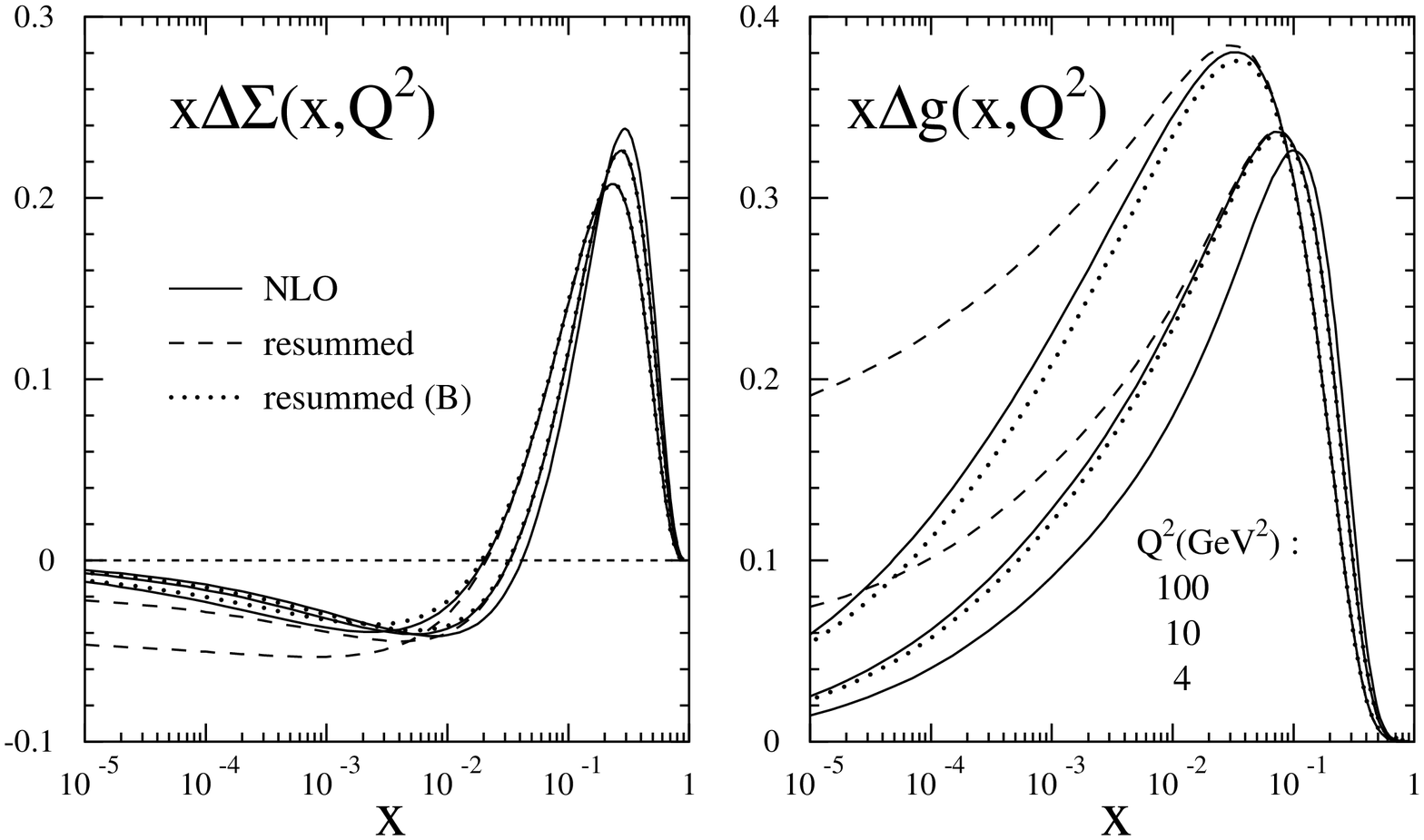,height=5.1cm,width=9.5cm}}
\vspace*{-4.2mm}
\end{center}
{\footnotesize
 \sf Figure~2:~The resummed evolution of the polarized gluon and singlet 
 densities as compared to the NLO results~\cite{RE9}. The possible impact
 of unknown less singular terms is illustrated by the curves (B), 
 cf.~eq.~(4).  The initial distributions at $Q^2_0 = 4 \mbox{ GeV}^2 $ 
 have been taken from ref.~\cite{REIN}.} 
\vspace{1.8mm}

\noindent
Numerical results for the evolution of $g_1^{\, ep,en}(x,Q^2)$ and the 
parton densities have been given for different input distributions~\cite
{RE9}. Fig.\ 2 shows an example. The situation is rather similar to the 
unpolarized case: the leading resummation effects are very large but 
unstable against less singular terms as $ x \rightarrow 0 $. 

\vspace{-2mm}
\subsection{QED non--singlet radiative corrections}
\noindent
The resummation of the $O(\alpha \ln^2 x)$ terms may yield 
non--negligible contributions to QED corrections. This has been shown
recently~\cite{RE11} for the case of initial state radiation for DIS
at large $y$. There the effect reaches around 10\% of the differential
Born cross section. The corresponding corrections to $\sigma(e^+e^- 
\rightarrow \mu^+ \mu^-)$ near the $Z$ peak are also discussed in 
ref.~\cite{RE11}.

\vspace{-2mm}
\section{Conclusions}
\vspace{-1mm}

\noindent
The resummations of the leading small-$x$ terms in both unpolarized and 
polarized, non--singlet and singlet anomalous dimensions have been 
investigated recently. At NLO the results agree with those found for the 
most singular terms as $x \rightarrow 0$ in fixed order calculations. 
Since the coefficient functions are known up to $O(\alpha_s^2)$ in the 
$\overline{\rm MS}$ scheme, predictions for the most singular terms of 
three--loop anomalous dimensions have been made~$^{2,6-8}$.

For non--singlet structure functions the corrections due to the 
$\alpha_s(\alpha_s \ln^2x)^{l}$ contributions are about $1\%$ or smaller 
in the kinematical ranges probed so far and possibly accessible at HERA 
including polarization~\cite{RE7,RE8}. 
The non--singlet QED corrections in deep-inelastic scattering resumming 
the $O(\alpha \ln^2 x)$ terms can reach values of about 10\% at $x 
\approx 10^{-4}$ and $y>0.9$~\cite{RE11}.

In the singlet case very large corrections are obtained for both 
unpolarized and polarized parton densities and structure functions~\cite
{RE9,RE6,RE10}. 
As in the non--singlet cases possible less singular terms in higher
order anomalous dimensions, however, which are in some cases required by 
conservation laws, are hardly suppressed against the presently resummed
leading terms in the evolution: even a full compensation of the 
resummation effects cannot be excluded.

To draw firm conclusions on the small-$x$ evolution of singlet structure 
functions also the next less singular terms have to be calculated. Since 
contributions even less singular than these ones may still cause relevant
corrections, it appears to be indispensable to compare the corresponding 
results to those of future fixed order three--loop calculations.
\vspace{2mm}

\noindent {\bf Acknowledgements :} This work was supported in part
by the EC Network `Human Capital and Mobility' under contract No.\
CHRX--CT923--0004 and by the German Federal Ministry for Research and 
Technology (BMBF) under contract No.\ 05 7WZ91P (0).

\end{document}